\documentclass[12pt,onecolumn,oneside,letterpaper,peerreview]{IEEEtran}

\usepackage{amsmath}
\usepackage{cite}
\usepackage{graphicx}
\usepackage{float}
\usepackage{times}
\usepackage{latexsym}
\usepackage{bm}
\usepackage{amssymb}
\usepackage[center]{caption2}
\usepackage{stfloats}
\usepackage{array}
\usepackage{fancyhdr}
\usepackage{cite,graphicx,amssymb}
\usepackage{citesort}
\usepackage{psfrag}
\usepackage{multirow}

\usepackage{color}

\ifCLASSINFOpdf

\else

\fi

    \def\Complex{{\rm\rule[.23ex]{.03em}{1.1ex}\kern-.3em{C}}}

    \newcommand{\be}{\begin{equation}} \newcommand{\ee}{\end{equation}}
    \newcommand{\bea}{\begin{eqnarray}} \newcommand{\eea}{\end{eqnarray}}
    \newcommand{\benum}{\begin{enumerate}} \newcommand{\eenum}{\end{enumerate}}

    \newcommand{\bl}[1]{\color{black}#1}

\makeatother

\begin{document}
\title{Massive Access for Future Wireless \\ Communication Systems}

\author{\IEEEauthorblockN{Yongpeng Wu, Xiqi Gao, Shidong Zhou, \\
Wei Yang, Yury Polyanskiy, and Giuseppe Caire}

\thanks{Y. Wu is with the Department of Electronic Engineering,
Shanghai Jiao Tong University, Minhang 200240, China (e-mail: yongpeng.wu@sjtu.edu.cn).
Y. Wu is also with State Key Laboratory of Integrated Services Networks,
Xidian University, Xi’an, China.}

\thanks{X. Q. Gao is with the National Mobile Communications Research Laboratory, Southeast University, Nanjing 210096,
P. R. China (e-mail: xqgao@seu.edu.cn)}

\thanks{S. Zhou is with the Department of Electronic Engineering, Tsinghua University, Beijing 100084, China
(e-mail: zhousd@tsinghua.edu.cn)}

\thanks{W. Yang is with Qualcomm Technologies, Inc.,
San Diego, CA, 92121 (e-mail: weiyang@qti.qualcomm.com).}

\thanks{Y. Polyanskiy is with the Department of Electrical Engineering and Computer Science,
Massachusetts Institute of Technology, 	77 Massachusetts Avenue Cambridge, MA 02139 (e-mail:yp@mit.edu)}

\thanks{G. Caire is with the Institute for Telecommunication Systems, Technical University Berlin, Einsteinufer 25,
10587 Berlin, Germany (Email: caire@tu-berlin.de).}

}
\maketitle

\begin{abstract}
Multiple access technology played an important role in wireless communication
in the last decades: it
increases the capacity of the channel and allows different users
to access the system simultaneously. However, the conventional multiple access technology,
as originally designed for current human-centric wireless networks,
is not scalable for future machine-centric wireless networks.

Massive access (studied in the literature under such names as ``massive-device multiple access",
``unsourced massive random access",
``massive connectivity", ``massive machine-type communication", and ``many-access channels")
exhibits a clean break with current networks
by potentially supporting millions of  devices in each cellular network.
The tremendous growth in the number of connected devices requires a fundamental rethinking of the conventional multiple access technologies in favor of new schemes suited
for massive random access. Among
the many new challenges arising in this setting, the most relevant are: the fundamental limits of
communication from a massive number of bursty devices transmitting simultaneously with short packets,
the design of low complexity and energy-efficient massive access coding and communication schemes, efficient methods for the detection
of a relatively small number of active users among a large number of potential user devices with sporadic transmission pattern,
and the integration of massive access with massive MIMO and other important wireless communication technologies.
This paper presents an overview of the concept of massive access
wireless communication and of the contemporary research on this important topic.

\end{abstract}

\section{Background: Multiple-Access with Small Number of Active Users}
In wireless communication systems, multiple access is a technique
to allow multiple users to obtain
the access to the system simultaneously using a shared communication media.
Multiple access technologies can be generally categorized
into coordinated multiple access and uncoordinated multiple access (also known as random access).

Coordinated scheme refers to a scenario where the users (i.e., the transmitters) are coordinated by a central unit (typically an access point or a base station)
prior to their transmissions. In contrast, uncoordinated access refer to scenarios where such coordination is absent. As we shall see,
such a classification depends critically on how we define coordination.
In this paper, we shall consider a multiple access communication scheme being coordinated
if each of the transmitters has some unique signaling/signature that is can be used by the receiver to perform detection.
We remark that, such a classification method is not perfect and sometimes can be subjective.
However, even if they are inaccurate, such classifications are needed to determine the important
factors for system modeling and for comparing different multiple access approaches.

\subsection{Coordinated Multiple Access}
Coordinated multiple access technology requires
dedicated
multiple access protocols to coordinate the communication
of the accessible users in the systems.
Typical coordinated multiple access technologies
can be summarized as follows.

\begin{enumerate}
\item Frequency division multiple access (FDMA) divides the bandwidth of the channel into separate
non-overlapping frequency subchannels,
 and assigns each subchannel to a separate user.

\item Time division multiple access (TDMA) uses
non-overlapping time slots to serve different users.

\item Code division multiple access (CDMA)
serves multiple users simultaneously
over the same frequency band by using different codes.
In CDMA, each user is allocated to the full available bandwidth
instead of transmitting over separate frequency sub-band.

\item Space-division multiple access (SDMA) utilizes
spatial beamforming technology to communicate
with multiple users in the same time at the same frequency.
The directional beamforming avoids interference among users.

\item Non-orthogonal multiple access (NOMA) allows more users
communicating simultaneously via non-orthogonal resources. The fundamental principle
behind NOMA technology is the classical multiuser information theory.

\end{enumerate}

The above coordinated multiple access technologies in
their conventional form have become mature,
and have been incorporated into various wireless broadband
standards. For example, the basic
FDMA technology was implemented in the first generation (1G)
cellular systems.
The second generation (2G) cellular systems were based on a combination of TDMA and FDMA,
and CDMA was the main access technology in the third generation (3G) cellular systems.
An advanced form of FDMA, orthogonal frequency-division multiple access (OFDMA) scheme is used
in both fourth generation (4G) and fifth generation (5G) cellular systems.
Also, SDMA in the form of multiuser MIMO is expected to be at the center of the forthcoming 5G.

As the telecom industry started the standardization of 5G technology,
several NOMA schemes were proposed to the 3rd Generation Partnership Project (3GPP)
as candidates for multiple access and were  thoroughly studied.
However, the number of users per degree of freedom (d.o.f)\footnote{per d.o.f means per signal dimension in time-frequency-antenna domain}
by current multiple access technologies is modest. For example, the typical overloading factors (the number of users per frequency domain)
considered in the most recent 3GPP study are between 1.5 and 3, beyond which the system performance may suffer significant degradation.

\subsection{Uncoordinated Multiple Access}
Uncoordinated multiple access
lets a set of users
to transmit over a common
wireless medium opportunistically and independently.
In practice, the reason not to perform a sufficient
coordination among the users trying
to access the system might be various,
e.g., the low latency requirement
to restrict the
coordination establishment,
a lack of global scheduling
information, or the bursty and random access pattern
for the activity of the users.
Uncoordinated multiple access technology  has been widely
used in the initial access process of both cellular terrestrial and satellite communication
networks \cite{Morlet2007SC,Ali2017CM}, where a random access mechanism is implemented.

It should be noted that
uncoordinated multiple access in definition of this paper is not equivalent to ``grant-free".
For coordinated multiple access,
the traffic pattern of each user (i.e., the user activity) does not need to be deterministic.
In other words, each user may communicate with the receiver randomly, and the active users in each communication
duration may be unknown at the receiver \emph{a priori}.  This communication scenario is typically referred to
as grant-free multiple access. However, the key difference
between a coordinated grant-free multiple access and an uncoordinated multiple access is that, in the former scenario,
each user is assigned with a dedicated pilot/signature sequence, which could be used by the receiver to identify its presence.
In contrast, in uncoordinated multiple access, the users all share the same transmission protocol,
no signature allocation/coordination is performed prior to each transmission. Thus, from the perspective of the receiver, the users are unsourced.

ALOHA is a classical solution for the unsourced random
access \cite{Roberts1975SCCR}, which follows a simple model in which
all transmissions with overlapping d.o.f. (``collisions'') are treated as complete
erasures, from which nothing can be recovered. Following a collision a retransmission is
attempted (by each of the participants of the collision), and a lot of research went into
ensuring that the resulting algorithms are stable (i.e., do not enter endless retransmission
loops).

While very simple to implement, a crucial bottleneck of the ALOHA is that only $1/e \approx 37\%$ of the d.o.f.
carry useful (uncollided) information. A recent approach to alleviate this limit is to
use forward error correction (preventive packet retransmissions) and the successive interference cancelation (SIC)
strategy to cancel replicas of
successfully decoded packets. This idea, known as coded-slotted-ALOHA (CS-ALOHA) \cite{Paolini2015TIT},
allows one to achieve almost $100\%$ efficiency.

In this paper, we focus on one-shot noninteractive multiple access scenarios, in which a user (i.e., the transmitter)
can only transmit but can not receive. There is a large body of literature on random access with
interaction or feedback (for example the carrier sensing multiple access (CSMA),
in which a user can hear whether other nodes are transmitting after a very small prorogation delay relative to a packet transmission time).
Interactive multiple access schemes are beyond the scope of this paper.

\section{Going Large: Massive Access}
{
All technologies reviewed in the previous section mostly break down when we
increase the number of users (per spectral d.o.f.). For example, for
coordinated multiple access, the overhead of coordinating this many users overwhelms the system.
The ALOHA results in too many collisions, effectively rendering communication impossible.
The CS-ALOHA requires many packet retransmissions dramatically increasing the effective energy-per-bit.
}

\begin{figure}[!t]
\centering
\includegraphics[width=0.8\textwidth]{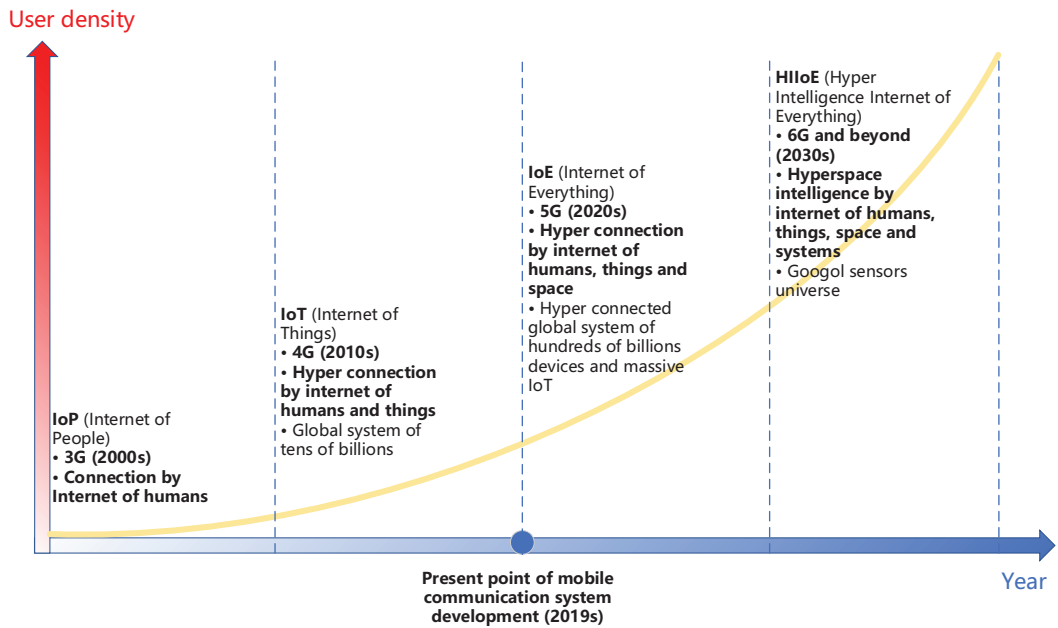}
\caption{Evolution of number of devices connected in the global communication systems.}
\label{IoE_Trend}
\end{figure}

Massive access is an emerging technology, that accommodates the number of users
per transmission medium by possibly orders of magnitude higher
compared to current state-of-the-art.
With massive access,  we refer to a scenario in which wireless networks
simultaneously serve millions of
infrequently communicating devices.
{This is driven by {\bl fast transition} from Internet of Things (IoT) wireless networks
towards future Internet of Everything (IoE) wireless networks, which includes the intelligent connection
of people, process, data and things.}
A Cisco research report reveals that $99.4\%$  of physical objects
that can be part of the IoE concept are not currently interconnected \cite{Bradley2013}.
Therefore, the number of accessible devices
in the wireless networks will increase dramatically in future.
Figure \ref{IoE_Trend} provides the evolution of number of devices
connected in the global communication systems
from the past Internet of People
(IoP, connecting people), to current IoT,
and finally future interconnection of everything as the concept of IoE \cite{5G_Forum}.

{
The difficulties with massive
access can be summarized by the slogan: ``Supporting
1M users at 10bps is much more difficult than 10 users
at 1Mbps \cite{Polyanskiy2018}".} This is because massive access has several significantly different properties.
The unique features of the massive access research
are as follows:

\begin{enumerate}

\item The number of active users in each d.o.f.
are orders of magnitude larger than the traditional model (i.e., massive).
 When the number of users goes to infinity,
a positive rate as in the classical multiple access channel
setup is not achievable \cite{Chen2017TIT}.
This is one of the main reasons why SIC cannot guarantee an {\bl arbitrarily small} error decoding
probability for massive access channel.

\item The flow for each device in massive access is often
expected to be a relatively small quantity of information to transmit (e.g., several hundred bits).
Explicit characterizations for the fundamental limits
of massive access communication with finite blocklengh and finite payload size
are still unknown. However, various upper and lower bounds are obtained in the literature.
These bounds indicate that the fundamental limits of massive access communication
are drastically different from traditional Shannon-type results, which we will
review in this paper.

\item To approach the fundamental limits,
new (and rather non-traditional) architectures are required.
We will review these new architectures in this paper.

\item For massive access in wireless fading channels,
it is unreasonable to assume that the receiver
knows the exact channel state information (CSI) of
the active users in advance, even for the statistical
CSI \cite{Fengler2019}. There are two reasons for this point. First,
for {a} massive number of devices,
assigning a dedicated pilot to each device for channel estimation is difficult.
More importantly, the
knowledge of the CSI for all potential devices requires
that each device's channel is calibrated and the attenuation due to the
propagation pathloss is stored in memory at the time of deployment.
Such calibration procedure must be repeated over time since the
propagation conditions change.
For millions of sensor devices,
maintaining such calibration is extremely intractable.
Even the statistical knowledge
of channel coefficients is hard to obtain in practice.
Extracting the statistical CSI requires collecting
a large sample
of the received signal energy from each device in the network.
This is tantamount to establish
a combined histogram for millions of sensor devices.
Similar to the calibration,
this needs a large amount of data exchange
which is difficult to implement in practice.

\end{enumerate}

\begin{figure}[!t]
\centering
\includegraphics[width=0.8\textwidth]{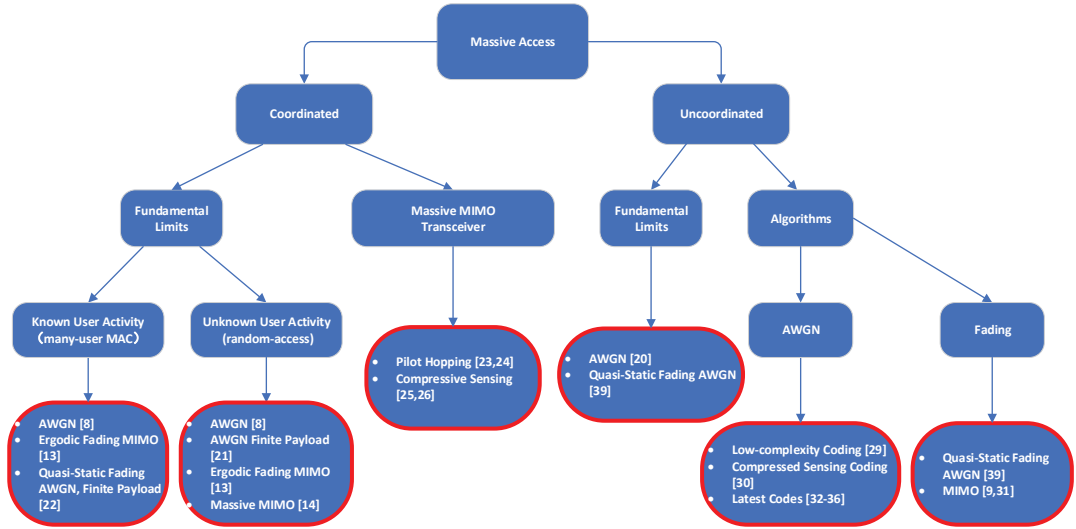}
\caption{The general framework for the recent typical work on massive access
research.}
\label{general_framework}
\end{figure}

To investigate massive access wireless communication,
the above-mentioned features have to be taken into consideration
jointly. {A summary of the recent work on this topic is sketched in Figure \ref{general_framework}.}
In general, the massive access research can be categorized
into two separate parts: coordinated massive access
and uncoordinated massive access.
In the following, we discuss the details of these work.

In this paper, we will discuss the fundamental
limits of massive access in both asymptotic and non-asymptotic regime.
The asymptotic regime is defined as $K_a \rightarrow \infty$, or $n \rightarrow \infty$, or $N_r \rightarrow \infty$,
where $K_a$, $n$ and  $N_r$ are the number of active users, {\bl the block length} and the number of receive antennas, respectively.
The non-asymptotic regime is defined as that $K_a$, $n$ and  $N_r$ are finite.
The non-asymptotic regime is more relevant from a practical point of view, but a good understanding of the fundamental limit in
the asymptotic regime may provide good insights into the behavior of complicated bounds and codes.
{\bl There are some previous review papers discussing conventional machine-to machine communications where the overloading
factor is less than 1 and orthogonal scheme is used \cite{Lien2011CM,Zheng2014WCM,Biral2015DCN}. In contrast, this paper focuses on
future massive machine type communications where overloading factor is much larger and non-orthogonal scheme is used.}

\section{Fundamental Limits for Coordinated Massive Access}
\subsection{The Infinite $E_b/N_0$ Regime}
Classical multiuser information theory establishes
the capacity region of conventional MAC with a fixed
number of users comparing to the asymptotically infinity
coding blocklength. However, when the number of active users
goes to infinity, there is a lot of multiuser interference (MUI).
As a result, the conventional information rate of Gaussian MAC
defined {by information bit per channel use for each user  approaches zero (i.e., $\frac{1}{K_a} {\log(1 + K_a P)} \mathop  \to \limits^{{K_a} \to \infty } 0$)
for coordinated massive access communication.}

To address this issue, Chen \textit{et al.} propose a concept
of symmetric message-length capacity for massive access
communication over additive white gaussian noise (AWGN)
channels \cite{Chen2017TIT}. The message-length capacity is
defined under three important assumptions:
1) Each user is assigned an individual codebook;
2) The number of user and the blocklength grow to infinite at a given speed;
3) The average error probability is defined as the probability
that the decoder mistakenly estimate
the message for any of the users.
Message-length capacity calculates
the information bit transmitted along the entire blocklength.
{Thus, although the conventional notion of information bit per channel use for each user goes to zero,
we can still send a total amount of information that is arbitrarily large.}
{In general, the message-length capacity is still an asymptotic analysis
which assumes infinite blocklength and infinite $E_b/N_0$ \cite{Polyanskiy2018}.
}

The model in \cite{Chen2017TIT} comes with two simplifications:
1) the gains from each user are the same;
2) the message-length rate for each user are the same as well.
These simplifications are relaxed in \cite{Wei2019},
where the authors consider multiple-input multiple-output (MIMO) with non-symmetric rate.
The message-length capacity region of the non-symmetric MIMO massive access
channel is derived in \cite{Wei2019}.
Two important observations are made: 1) When the number of receive antenna is finite,
similar to the AWGN case, the authors show
that the conventional notion of capacity region is not meaningful.  Instead, they reveal
that message-length capacity region of MIMO massive access channel
is dominated by sum rate constraint only, and the individual user rate is determined by a
specific factor that corresponds to the allocation of the sum rate.
Moreover, it is shown that asymptotically, there is no benefit of having multiple antenna
at the transmitters for the individual rate and what matters is the number of antennas
at the receiver; 2) When the number of receive antennas grow to infinity,
SIC is able to work and the capacity region reduces
to that of the conventional MIMO MAC. The basic intuition behind
this result is that the receive power per degree of freedom goes to infinity
when the number of receive
antenna grows unbounded.
As a result, the error probability of each user decays
exponentially and a positive per
channel use rate can be achieved.

\begin{figure}[!t]
\centering
\includegraphics[width=0.8\textwidth]{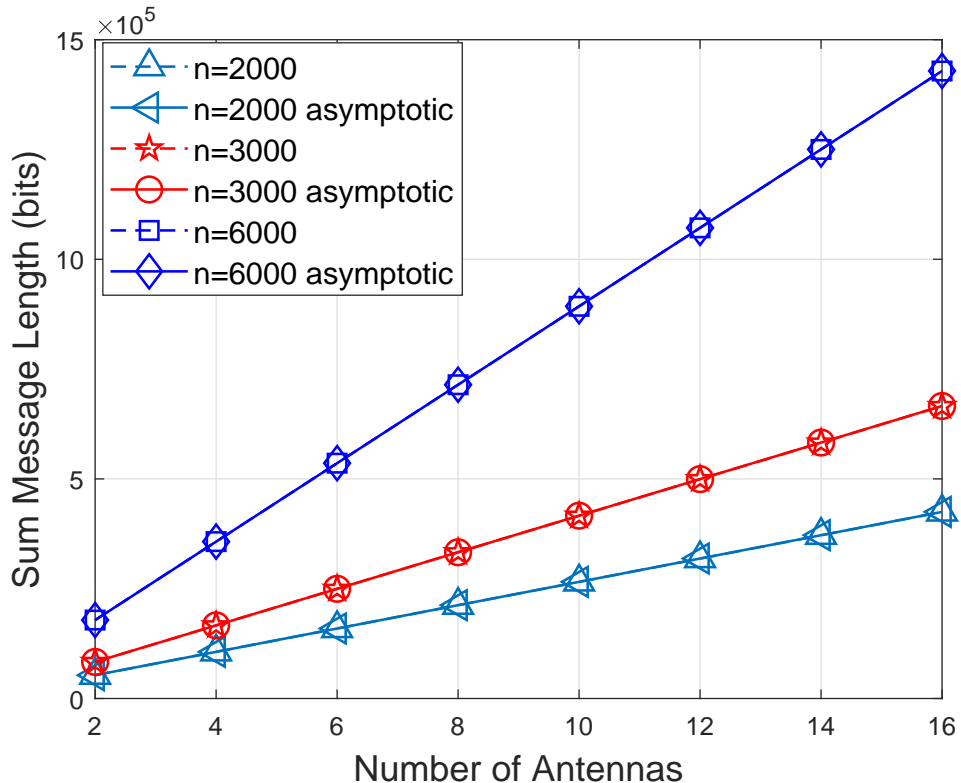}
\caption{The capacity of MIMO massive access versus growing receive
antennas.}
\label{Sum_rate_massive_access}
\end{figure}

By considering
a single cell communication system with $k_{n}$ users randomly deployed within the cellular network,
Figure \ref{Sum_rate_massive_access} shows
the impact of receive antenna numbers $N_r$
on the capacity of MIMO massive access communication.
Both the asymptotic and the exact sum-rate expressions are given in \cite{Wei2019}.
Each user has a transmit power $P_{k}$ generated uniformly within the interval $[5,15]$.
The channel of each user conforms to i.i.d. Rayleigh fading.
To meet the massive connectivity requirement,
the number of users is set to $k_{n}=n/2$, which grows linearly with the codelength $n$.
Each user is equipped with $N_t = 4$ transmit antennas.
We observe
from Figure \ref{Sum_rate_massive_access}  that given different codelength $n$, the capacities
all grow linearly with an increasing number of $N_r$.
This is because the number of degrees of freedom of a MIMO system is
limited by the minimum of the number of transmit antenna
and the number of receive antenna.
For massive access communication,
the ratio of the sum signal power to noise power is high\footnote{The sum capacity depends only on the sum signal power.}.
Therefore, we have $N_{\mathrm{DoF}} = \min\{N_{r},k_{n}N_{t}\}=N_{r}$.
As a result, the capacity of MIMO massive access
grows linearly with the number of receive antenna  as shown
in Figure \ref{Sum_rate_massive_access}.

Finally, as discussed in Section II and illustrated in Figure \ref{general_framework},
user identification is an important step in coordinated massive access
communication. The authors in \cite{Chen2017TIT} further investigate the model
where only an unknown subset of users send data.
These active users utilize unique signatures
to enable their identification by the receiver.
The capacity penalty due to this
user activity uncertainty
is obtained as a form of the identified signature length.
One important message from \cite{Chen2017TIT}
is that, asymptotically when number of users and blocklength
grow to infinity with a fixed rate, and
the payload of each user grows to infinity
logarithmically with the number of active user,
it is {\bl asymptotically}  optimal
to do a two-step approach: a step of active user detection,
and a communication stage (see also \cite{Wei2019} for a generalization to the MIMO case).
{When the payload of each user is finite, using a partial resource
to perform activity detection separately can
be suboptimal.  However, the optimal communication scheme
is still unknown. For MIMO case when $K_a/N_r = o(1)$,
another observation for active user detection is that we can identify the activity of $K_a = O(D_c^2/\log^2(K/K_a))$
active users \cite{Haghighatshoar2018}, where $D_c$ and $K$ represents a coherence block length and the total
number of users, respectively. This is much larger than
the previous bound $K_a = O(D_c/\log(K/K_a))$ obtained via traditional compressed
sensing techniques \cite{Kueng2018TIT}.}

{
\subsection{The Finite $E_b/N_0$ Regime}
For practice, the
infinite packet-size/infinite $E_b/N_0$ asymptotics are not very relevant. In order to fix this
problem, we need to abandon the traditional joint probability of error criterion (dominant in the multiuser information theory literature since
the days of Liao, Ahslwede, Cover and Wyner \cite{Liao1972ISIT,Ahlswede1971ISIT,Cover1975book,Wyner1974TIT}) and introduce per-user probability of error (PUPE)~\cite{Polyanskiy2017_2}.
More specifically, PUPE is defined as the average (over the active users) fraction of the transmitted messages that are not  decoded correctly.

Under PUPE \cite{Polyanskiy2017_2} and most recently \cite{Zadik2019ISIT} showed progressively tighter bounds on the fundamental
energy-per-bit required for reliable massive access communication. More exactly, those work consider the
regime where the number of bits transmitted by each user (``payload'') is fixed, and the number of users
grows to infinity linearly with the blocklength (so that the density of users per d.o.f.,
and hence the spectral efficiency, is held constant).
The two key observations of \cite{Polyanskiy2017_2,Zadik2019ISIT} are the following:

\begin{enumerate}
	\item There exist coded-access architectures which exhibit perfect multi-user interference
cancellation at low spectral efficiencies. For example, for payload of 100 bits and spectral efficiencies
below 1 bps/Hz, it is possible to arrange communication of arbitrarily large number of users so that each one's
energy-per-bit is almost exactly the same as if he was communicating alone (without any MUI). We
call this a perfect MUI-cancellation property.

\item The orthogonal massive access (TDMA, FDMA, CDMA) do not have the perfect MUI-cancellation property.
In other words, even in the coordinated massive access setting, orthogonalizing access is severely
suboptimal from the energy-efficiency point of view.

\end{enumerate}

Kowshik and Polyanskiy study the same regime, but consider the effects of Rayleigh fading \cite{Kowshik2019}.
Same behavior is observed
(at low user density regime, perfect MUI cancellation is possible).
Also, it is observed that a larger energy-per-bit is required for reliable massive access
over the quasi-static Rayleigh fading channel compared to the AWGN case to overcome the randomness in the fading gains,
especially when CSI is not known at the receiver. However, this randomness is shown to be helpful for
the decoder to differentiate users.

}

\section{Coordinated Massive Access Meets Massive MIMO}
Massive MIMO technology, which utilizes a very large number of antennas at
the BS with simple signal processing to provide
services for a comparatively small (compared to the number of antennas)
number of active devices, is regarded as a key
technology for future wireless communication systems.
By providing spatial resolution within the same time/frequency
resource, massive MIMO system can support massive access communication
in a better way.

Carvalho \textit{et al.} propose a joint pilot assignment and data
transmission protocol for crowded massive MIMO system,
where each user selects
a pilot signal in each time slot based on a pseudo-random pilot
hopping pattern known at the BS \cite{Carvalho2017TWC}.
Therefore, over an asymptotically long time horizon,
the BS can detect the active users based on the specific pilot
hopping pattern. Moreover, the effects of pilot signals collision
can be averaged out to enable a reliable rate for transmission.
Gao \textit{et al.} extended the  pseudo-random pilot
hopping scheme to practical spatially correlated
massive MIMO channels \cite{GaoGlobecom2019}.

A key characteristic of the
massive access in future wireless network lies in the sporadic traffic of users, i.e., only a small number
of users are activated to access the network in any given time interval.
By exploiting this sporadic traffic, a compressive sensing-based
scheme is proposed to detect user activity and estimate the channels for massive
access in massive MIMO systems \cite{Liu2018TSP}. The proposed scheme is
based on the classical AMP framework from
compressed sensing.
It is proved in \cite{Liu2018TSP}
that by utilizing AMP technique to exploit sporadic traffic
property of massive access, the missed user detection probability and the
false alarm probability for activity detection can
be suppressed to zero in the asymptotic antenna dimension regime.

Ke \textit{et al.} propose several compressive sensing-based adaptive active user detection
and channel estimation schemes by exploiting the
virtual angular domain sparsity of more practical massive MIMO channels \cite{Ke2019}.
Here we provide some simulation results for the massive access in massive MIMO systems in \cite{Ke2019}.
In the simulations, the BS employs uniform linear array of $M =64$ antennas,
$K = 500$ potential users
are randomly distributed in the cell with radius 1 km, and $K_a = 50$ (${K_a} \ll K$) users are active.
The results here are obtained by averaging over $N_{\rm{sim}} = 1000$ numerical simulations.
The system adopts OFDM for the massive access in enhanced mobile broadband scenario,
where $N = 2048$ subcarriers and the cyclic prefix of the length $N_{\rm{CP}}=64$ are considered.
The length of pilot signals is set to be the same with the length of cyclic prefix
for the estimation of the maximum multipath delay $N_{\rm{CP}}$.
For each user's massive MIMO channel, we consider the one-ring channel model.
Each user is with the limited angle spread seen from the BS, so that each user's
massive MIMO channel exhibits clustered sparsity in  virtual angular domain.
We assume this effective angular-domain sparsity level $S_a$ varies from 4 to 10.

Three schemes are evaluated for the above-mentioned massive-access setup {\bl \cite{Ke2019}}:
\begin{itemize}
\item{\textbf{Scheme 1}:
The spatial-frequency structured sparsity is exploited, where both channel gain-based activity detector (CG-AD)
and belief indicator-based activity detector (BI-AD) are used. }

\item{\textbf{Scheme 2}:
Both spatial-frequency and angular-frequency structured sparsity are simultaneously exploited,
where both CG-AD and BI-AD are used.}

\item{\textbf{Scheme 3}:
Apply Turbo-structure, and where spatial-frequency and angular-frequency structured sparsity are alternately exploited,
where only BI-AD can be used.}

\end{itemize}

{\bl
Scheme 1, Scheme 2, and Scheme 3 are described in details in \cite[Section III-B-1)]{Ke2019},
\cite[Section III-B-2)]{Ke2019}, and  \cite[Section III-C-1)]{Ke2019}, respectively.
}

\begin{figure}[!t]
\centering
\includegraphics[width=0.8\textwidth]{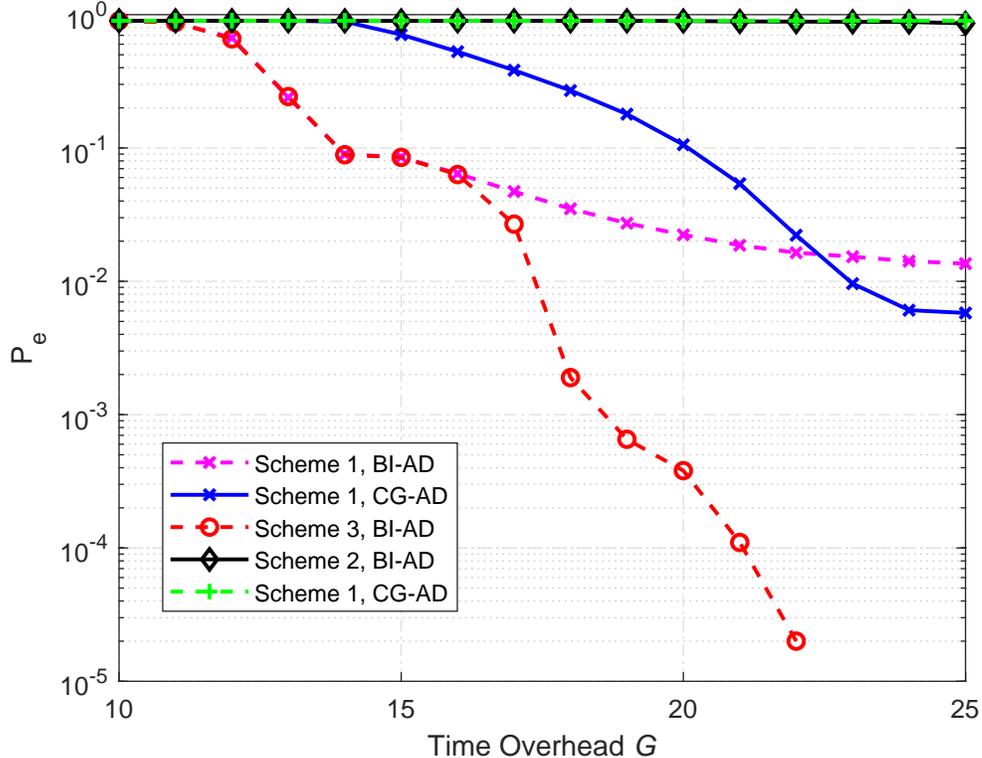}
\caption{Comparison of  error detection probability  of Schemes 1--3 as functions of $G$.}
\label{Pe_result}
\end{figure}

We observe from Figure \ref{Pe_result} that \emph{Scheme 3}
using the iteration structure can achieve  much better active user detection and channel estimation
performance\footnote{For error detection probability in Figure \ref{Pe_result}, we count an error if any user is mis-detected or false alarmed.} than \emph{Scheme 1} and \emph{Scheme 2} even for the very low training overhead $G \ll K_a$,
 which shows its superiority in the considerable reduction of access latency.

{\bl \section{Uncoordinated Massive Access}}
Unsourced random access mechanism is implemented for
the initial access in both cellular terrestrial and satellite communication
networks \cite{Morlet2007SC,Ali2017CM}. In contrast to coordinated
multiple access protocols, a number of uncoordinated
users who share a common channel (codebook) send their data opportunistically
and retransmit when collision occurs.
Several multiple access
coding schemes are proposed in the literature
to address the unsourced random access problem, such as slotted ALOHA  \cite{Paolini2015CM,Paolini2015TIT}
and TIN \cite{Polyanskiy2017_2}.
However, for massive random access when the set of active users
become large, these schemes
are not able to work efficiently anymore \cite{Polyanskiy2017_2}.

{
R. Gallager appeals to ``a coding technology that is applicable
for a large set of transmitters of which a small,
but variable, subset simultaneously use the channel"
more than 30 years ago \cite{Gallager1985TIT}.
To the best of our knowledge, this coding scheme
has not been perfectly designed so far.
To shed light on the fundamental limitations of this problem,
Y. Polyanskiy proposed a novel formulation
for massive random access \cite{Polyanskiy2017,Polyanskiy2017_2}.
Comparing to the conventional coding schemes for coordinated massive access,
Polyanskiy's framework is developed under three
key  assumptions: 1) All the users share a common codebook;
under the common codebook assumption, the decoder only {\bl needs} to decode
a list of messages transmitted from the active users, but the identities of the set of users do not need to be recovered\footnote{In some applications, the identity of the source may be needed by the receiver. For example, specific sensors need to update their status periodically. To apply this framework to these applications, each user could embed its identity inside its payload. However, we remark that the user identity is recovered at the higher layers of the communication protocol (by interpreting the content of the payload) rather than the physical layer.}. 2) The error probability is defined as the average fraction of mis-decoded messages
over the number of active users, which is referred to as PUPE.
This is also a more reasonable performance evaluation metric
since the requirement of correctly decoding for every user's message
as in the classical multiuser information theory is intractable
when there are ``infinity many" users in the system.  Moreover,
from a practical perspective, a user typically does not care about whether other users' messages are correctly decoded or not.
 3) Each user transmits a fixed small amount of information bits  within
a finite code length $n$.  The analysis under
finite blocklength and finite payload assumption in \cite{Polyanskiy2017_2}
indicate that the minimal energy per bit required for a given error
probability has a certain ``inertia":
as the user density $\mu $ increases from zero, initially the energy-per-bit stays
the same as in the single-user case. This implies that the optimal multiuser access architectures should be able to perfectly
cancel all multiuser interference and achieve an essentially single-user performance for each user as long as the user density is
below a critical threshold\footnote{Interestingly, the same effect is observed
for coordinated massive access scenarios in Section III, and for both
AWGN and fading channels (see also \cite{Polyanskiy2018,Kowshik2019,Zadik2019ISIT}).}. Note that this is significantly
 different from traditional Shannon-type multiuser results based on
 asymptotic analysis. }

\begin{figure}[!t]
\centering
\includegraphics[width=1\textwidth]{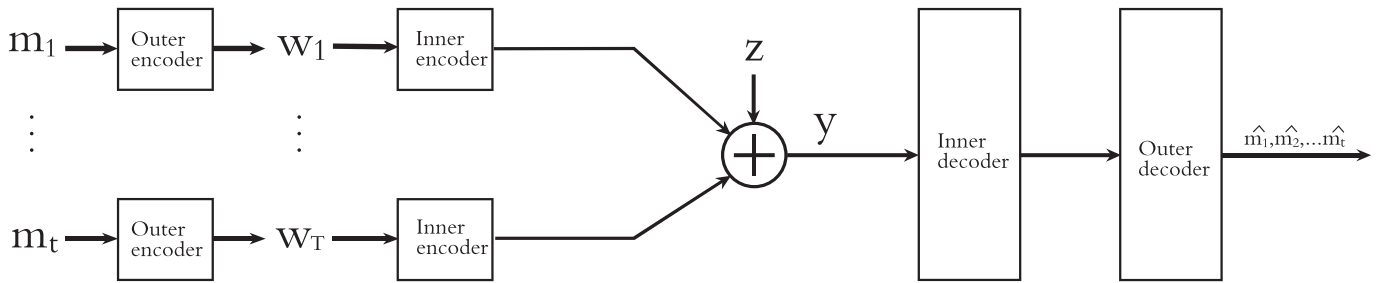}
\caption{Concatenation coding structure under Polyanskiy's massive unsourced
random access framework.}
\label{Yury_code}
\end{figure}

The above-mentioned three aspects formulate a new
information theory framework for massive access
communication. Under Polyanskiy's massive unsourced
random access framework, it is revealed
in \cite{Polyanskiy2017_2} that existing schemes such as
``treat noise as interference" (TIN) and
ALOHA are very far from the analytical
random coding bounds. To improve the performance,
a $T$-fold ALOHA approach is proposed in \cite{Polyanskiy2017},
where the $n$ channel uses are divided
into $V$ sub-blocks (i.e., slots), and each active user randomly
select one sub-block.
If the number of users in each sub-block is
no more than $T$, then the decoder try to decode all corresponding
messages; otherwise, nothing is decoded.
To implement $T$-fold ALOHA in practice, a low complexity
coding scheme for the random access channel with $T$ active users
is also proposed in  \cite{Polyanskiy2017}.
In the proposed scheme, each user encodes its message based on
a common concatenation code, as illustrated in Figure \ref{Yury_code}.
The inner code is a linear code, which converts $T$-user Gaussian MAC
into a mod $p$ (noiseless) adder MAC\footnote{For the coding scheme given in \cite{Polyanskiy2017}, $p =2$.
Therefore, the sum rate of the scheme in \cite{Polyanskiy2017} is less than 1 bit/channel use.}.
The outer code encodes
the individual message for each user
that embedded in the mod $p$ sum.

Another principal
discovery of \cite{Polyanskiy2017_2} is the following. The main practical goal is to study
finite-blocklength performance bounds (for example, minimal $E_b/N_0$ for
a fixed number of degrees of freedom $n$, fixed PUPE $\epsilon$, fixed
payload $k$ and variable number of active users $K_a$). However, the
resulting bounds are quite cumbersome and do not admit easy analytical
interpretation. Fortunately, the finite blocklength questions at
relevant values of ($n$, $k$, $K_a$) seem to be closely approximated by the
asymptotic question where $n\to \infty$, $K_a = \mu n$ and the payload
of each user is set to $M=2^k K_a$. The effective $E_b/N_0$ should be
defined as energy of the codeword normalized by $2k$ (instead of $2
\log_2 M$). Studying asymptotics in this regime appears to closely
resemble behavior of the fundamental limits at finite $n$.

Following Polyanskiy's massive unsourced
random access framework, Amalladinne \textit{et al.} propose a CS-based
coding scheme for AWGN channel by using a CS
code and a tree code as  the inner code and outer code in Figure \ref{Yury_code}, respectively \cite{Amalladinne2018}.
Each user's message is divided into smaller sub-message
and encoded by one column of a given coding matrix.
Parity bits are added to individual sub-messages
using a systematic linear block code.
The main purpose of the redundancy in \cite{Amalladinne2018}
is to allow the receiver to stitch together
all the sub-messages into a single long message.
Therefore, the parity bits are like cyclic redundancy check
bits that verify the connection between messages in two subsequent blocks.
The decoder first identifies which columns of the coding matrix have been transmitted
from the received noisy superposition.
This is a classical
``sparse support identification" CS problem and
can be efficiently solved based on standard algorithms.
Then, the decoder pieces together the individual segments
of the original messages where each sub-message sequence
represents a valid path in the tree.

Fengler \textit{et al.}  extend Polyanskiy's framework to multiple-antenna
channel \cite{Fengler2019}. A key difference between the AWGN channel and the multiple-antenna
fading channel is  whether channel estimation is required for detection.
For AWGN channel, the channel
coefficients of all users are regarded as 1 and the detection is implicitly
coherent by default. For multiple-antenna fading channel, it
may be  unreasonable to assume that all users' channel knowledge are known
a priori.  In standard multiuser MIMO communication systems,
dedicated pilot signal is allocated to
each user to allow the BS receiver
 estimate its channel knowledge and perform a coherent detection.
Apparently, assigning a dedicated pilot to each user does not
conform to the single common codebook framework.
Moreover, estimating the channel knowledge
of all users for massive access communication is
difficult. Based on the concatenated code structure,
Fengler  \textit{et al.} propose a
non-Bayesian inter code to perform
activity detection by treating the users' large-scale pathloss coefficients
as a deterministic unknown sparse non-negative vector.
Then an outer tree code as in \cite{Amalladinne2018} is used to
decode individual message for each active user.
Surprisingly, simulation results indicate that the proposed
scheme in \cite{Fengler2019}  outperforms the Bayesian vector
approximated message passing (AMP) even when
complete instantaneous CSI of each user is available at the receiver.
Ding \textit{et al.} incorporate active
user identification, channel estimation, and data decoding in a single phase \cite{Ding2018}.
To further improve the performance, a turbo AMP structure is proposed,
where the overall graphical model of the detection problem is divided into two
subgraphs: one represents the bilinear constraints of the channel model, and another represents
the structured sparsity of the user signals.
The two subgraphs exchange information iteratively until
convergence.

Polyanskiy's framework has attracted a wide research interest from different aspects recently. {\bl For AWGN
channel, Calderbank et al. \cite{Calderbank2018}, Fengler et al. \cite{Fengler2019_2},  Pradhan et al. \cite{Pradhan2018}, and
and Marshakov et al. \cite{Pradhan2019arxiv,Marshakov2019} use binary chirp code, sparse
regression code, short blocklength LDPC code, and Polar code as inter code in Figure 5 to obtain additional coding
gains, respectively.} Effros et al. develop an identity-blind decoding receiver architecture when the receiver
does not know the number of active user \cite{Effros2018ISIT}. Inan et al. study the active user detection problem by
exploiting a novel sparse group testing method \cite{Inan2017}. Kowshik et al. extend Polyanskiy’s framework into
quasi-static fading AWGN channel \cite{Kowshik2019arxiv}. It is revealed in \cite{Kowshik2019arxiv} that by
leveraging the
inherent randomization introduced by the channel allows for an easier
user separation and even low-complexity schemes can closely approach
optimality.

\begin{figure}[!t]
\centering
\includegraphics[width=0.8\textwidth]{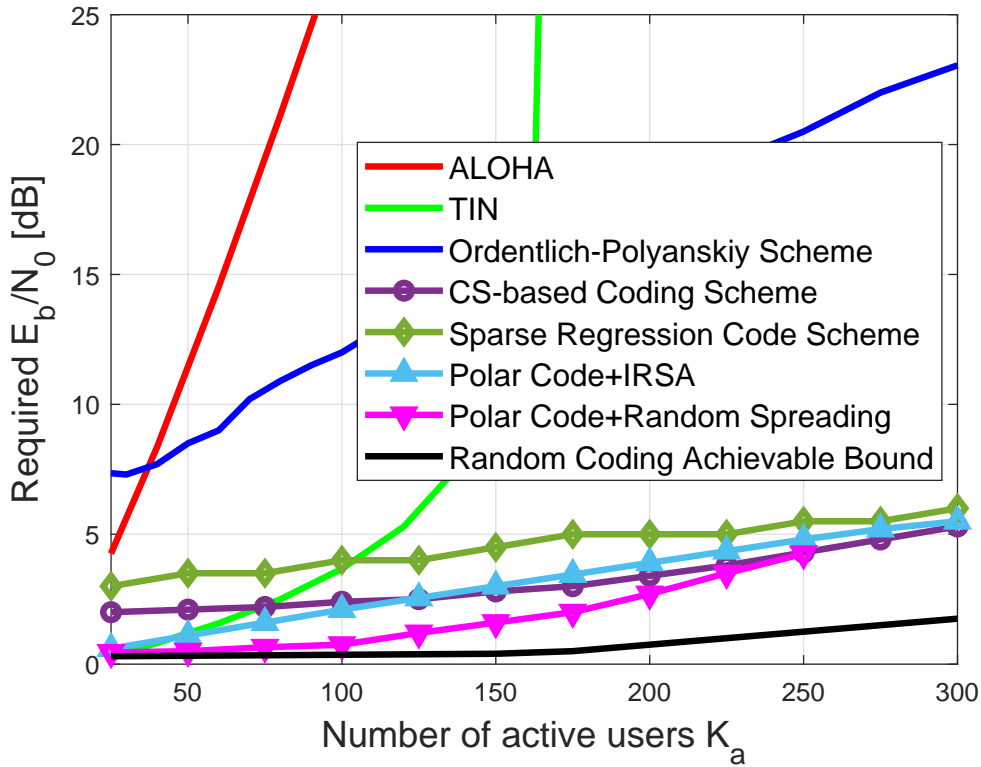}
\caption{Comparison between the $E_b/N_0$ required by various schemes for $k = 100$ bits per user,
$n = 30000$ channel uses, $P_e = 0.05$ error probability,
and different number of active users $K_a$}
\label{code_result}
\end{figure}

{\bl Figure \ref{code_result} plots the $E_b/N_0$ required by Ordentlich-Polyanskiy scheme \cite{Polyanskiy2017},
sparse regression code scheme \cite{Fengler2019_2}, CS-based coding scheme  \cite{Pradhan2018}, Polar code schemes \cite{Pradhan2019arxiv,Marshakov2019},
and relevant benchmarks for $k = 100$ bits per user,
$n = 30000$ channel uses, $P_e = 0.05$ error probability,
and different number of active users $K_a$.}
We observe from Figure \ref{code_result} that the
TIN scheme and the traditional ALOHA scheme
result in very poor energy efficiency in the massive access regime
whereas Ordentlich-Polyanskiy scheme is still effective
when $K_a = 300$.  However, we also observe from Figure \ref{code_result} that
Ordentlich-Polyanskiy scheme, CS-based coding scheme,
and sparse regression code scheme still have obvious gap comparing to an achievable
theoretical bound at $K_a = 300$. Therefore, the coding scheme has substantial space
to be improved.

{To summarize, current coding schemes for massive unsourced random access mainly revolve around
two ingredients: 1) exploit sparsity among user codewords to reduce collisions and interferences, where ALOHA or coded
slotted ALOHA technologies can be employed, 2) chunk the set of information bits into smaller-sized messages,
and use dedicated approach to decode each of the smaller-sized messages (e.g., compressive sensing \cite{Amalladinne2018}).
}

\section{Massive Access: Promising Research Potential}
Massive access reveals many  entirely new problems in
communication theory which needs research:
{\bl

\begin{itemize}

\item \textbf{Communication Scheme for Common Codebook}:
Massive access results in vast
amounts of active users
that need to be served simultaneously.
A simple access protocol is vital.
Therefore, communication scheme
for common codebook without additional
coordination is important.
The capacity-achieving coding for massive access communication
under the common codebook assumption is  missing.
The current coding scheme still has an obvious gap comparing to
the theoretical bound.
A sparse source coding compression and recovery approach
might be helpful for further improving the performance.

\item \textbf{User Identification}: A key characteristic of the massive
access is that there are a massive number of infrequently communicating users. In any
given instant, only a fraction number of users (still large) are active and
the activity pattern is unknown at the receiver and needs to be identified.
Therefore, user identification poses an important role
in massive access communication.
The receiver has to process a large amount of data from massive users in real time.
Some technologies from compressive sensing might be helpful.
Still, there are many issues to be further studied, e.g.,
user identification method in practice massive MIMO channels,
low processing complexity (linear or nearly linear) algorithms,
asynchronous user identification schemes,
and non-Bayesian user identification methods, etc.

\item \textbf{Imperfect CSI assumption}: For massive access in multiple antenna systems,
acquiring the exact instantaneous/statistical CSI of all users
is difficult and may cause a huge overhead.
As a result, imperfect CSI at the transmitters and the receiver
is a more reasonable assumption for practice massive access systems.
Robust design based on imperfect CSI
should be investigated.
This will result in new optimization problems and there
is enormous potential for the design of optimized algorithms.

\item \textbf{Short-packet Communication}:
For massive access communication, each user
sends a fixed number of bit to the receiver
due to finite energy-per-bit and low latency requirement.
The short packet communication will
cause a finite-blocklength effect.
For a large number of users, this will
result in a significant different
transmission strategy.
As a result, finite-blocklength analysis
is important for
massive access communication.
New system performance limits should
be derived via finite-blocklength
information theory
and new communication
schemes should be designed
based on finite-blocklength information rate.

\item \textbf{Energy-efficiency and Cost-efficiency}:
Massive access is a wireless technology to provide connectivity
to a massive number of low-power, low-cost, low-complexity devices.
Therefore, energy-efficiency and low-complexity signal
processing algorithms and hardware structures
are essential for massive access systems.
To realize finite energy-per-bit requires total power constraint
over the entire blocklength, finite payload of each user,
and PUPE constraint. The fundamental analysis
for these models reveal many new theoretical results which
are significant different from conventional Shannon-type results.
To approach the fundamental limits,
new architectures with low implementation complexity
are necessary.  Artificial intelligence aided communication
might be helpful.

\item \textbf{Privacy and Security}: Small payloads, low-latency constraints, and limited computational
capabilities of the devices make it hard to employ current cryptographic methods
in massive access wireless communication. Novel security and privacy protocols
need to be developed. Physical layer security has recently been recognized as a promising mechanism to achieve confidentiality
by exploiting the inherent randomness of wireless channels at the physical layer.
In particular, physical layer security can enable secure communication over the wireless medium without the aid of an encryption key.
This advantage makes physical layer security particularly suitable for implementation in networks with massive devices and heterogeneous subsystems.
As per requirements on physical layer security, no limitations are imposed on the eavesdroppers in terms of their computational capabilities.
This indicates that secure communications can still be achieved even if the eavesdroppers in futuristic networks are powerful and computational devices.
Therefore, physical layer security, operating essentially independently of higher layers,
is expected to augment the existing security mechanisms for safeguarding future
massive communication systems.

\item \textbf{Heterogeneity}: One challenge in developing general massive access
 solutions is that the massive number of devices
 might be extremely different in terms of
computational capabilities, cost, energy consumption, and transmission power.
As a result,  technologies which can address this heterogeneity
should be developed, which require  high efficiency
resource allocation, user scheduling,
and interference management schemes.

\end{itemize}

}

\section{Conclusions}
In this paper, we have  highlighted the significance
of massive access as a key enabling technology
for future beyond 5G wireless networks.
However, there are
still fundamental challenges ahead  for the practical implementation of massive access wireless communication,
e.g., when it comes to common codebook coding, realization of low complexity processing algorithms,
sporadic and small payload traffic of users, and synchronization protocols.
This provides researchers both in academia and industry a promising research potential
and many new research problems are revealed.



\end{document}